# Resource Allocation Mechanism for Media Handling Services in Cloud Multimedia Conferencing

Abbas Soltanian[†], Diala Naboulsi[†], Roch Glitho[†], Halima Elbiaze[‡]
[†]Concordia University, [‡]Université du Québec À Montréal
{ab_solta, d_naboul, glitho}@encs.concordia.ca, elbiaze.halima@uqam.ca

*Abstract*—Multimedia conferencing is the conversational exchange of multimedia content between multiple parties. It has a wide range of applications (e.g., Massively Multiplayer Online Games (MMOGs) and distance learning). Media handling services (e.g., video mixing, transcoding, and compressing) are critical to multimedia conferencing. However, efficient resource usage and scalability still remain important challenges. Unfortunately, the cloud-based approaches proposed so far have several deficiencies in terms of efficiency in resource usage and scaling, while meeting Quality of Service (QoS) requirements. This paper proposes a solution which optimizes resource allocation and scales in terms of the number of participants while guaranteeing QoS. Moreover, our solution composes different media handling services to support the participants' demands. We formulate the resource allocation problem mathematically as an Integer Linear Programming (ILP) problem and design a heuristic for it. We evaluate our proposed solution for different numbers of participants and different participants' geographical distributions. Simulation results show that our resource allocation mechanism can compose the media handling services and allocate the required resources in an optimal manner while honoring the QoS in terms of end-to-end delay.

## I. INTRODUCTION

Multimedia conferencing (or conferencing in short) can be defined as the conversational and real-time exchange of multimedia content (e.g., voice, video, and text) between several parties [1]. It has three main services: signaling, media handling, and conference control [2]. Media handling services offer different functionalities such as audio and video mixing, compressing, and transcoding. Several conferencing applications such as distance learning, video conferencing, and Massively Multiplayer Online Games (MMOGs) use audio extensively. They also use video, although to a lesser extent. Strategy MMOGs do use live video streaming for missions coordination [3]. Therefore, media handling services are of critical importance in conferencing applications. Moreover, in some conferencing applications like MMOGs [3], there might be thousands or hundreds of thousands of end-users (i.e., conference participants) who are geographically distributed. Thus, scalability is crucial in media handling services. Furthermore, the pressure of cost reduction brings the need for efficient use of resources. In addition, participants QoS requirements (e.g., end-to-end delay) need to be met.

Cloud computing [4] is an emerging paradigm where resources (e.g., storage, network, and services) are provisioned rapidly and on demand. It has several inherent benefits such as scalability and elasticity. These characteristics make it suitable for provisioning conferencing applications. This paper focuses on media handling services in cloud-based multimedia conferencing.

Fig. 1 depicts the assumed business model. It has four main roles: conferencing application providers, conferencing service providers, media handling service providers, and conferencing IaaS providers. In this model, conferencing applications rely on a conferencing service that is offered as a Software-as-a-Service (SaaS). Media handling services are also offered to conferencing service providers as SaaSs. The actual resources (e.g., CPU, RAM, and storage) for media handling services are provided by geographically distributed Infrastructure-as-a-Services (IaaSs).

In this paper, we propose a Cloud-based Resource Allocation Mechanism for media handling services in multimedia conferencing (CRAM). Our proposed solution optimizes resource allocation and scales in terms of the number of participants while guaranteeing QoS. As it is shown in Fig. 1, CRAM runs in the IaaS. CRAM allocates or deallocates resources for media handling services based on the fluctuation in the number of participants. It performs a fine-grained scaling of resources to improve efficiency in resource utilization. Moreover, to reduce the network cost and latency, it selects adequate locations for allocating resources. Besides efficient resource and network utilization, it caters to the QoS, with respect to media handling response time and network latency.

This paper is an extension of our previous cloud-based resource allocation mechanism [5]. Our proposed mechanism in [5] considered conferencing applications which need video mixing. Functionalities of Media handling services, such as audio and video mixing and compressing, are of critical importance in conferencing applications. In this paper, we take into account both mixing and compressing for media handling instead of mixing functionality only. We analyze the impact of participants' geographical distribution on the required

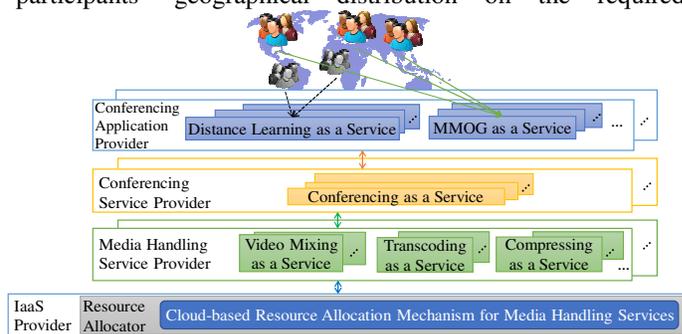

Fig. 1. Cloud-based conferencing business model

*This paper is an extended version of the paper published in CNSM2015, under the title "A Resource Allocation Mechanism for Video Mixing as a Cloud Computing Service in Multimedia Conferencing Applications"*

compression rate for video streams. In addition, we take into account the servers' resource cost and the network cost in the objective function of our ILP formulation while these were not considered in [5]. Indeed, the cost has a critical impact of the number of participants and their geographical distributions when allocating the resource for media handling services in Cloud-based multimedia conferencing. Thus, CRAM heuristic takes into account the cost parameter as minimization objective while meeting the QoS. Furthermore, instead of predefined zones, CRAM implements *DSort* algorithm to return the list of selected servers for a given set of participants' locations and servers' locations. Finally, we have considered extensive evaluation scenarios for evaluating the performance of our proposed CRAM solution, namely MMOG and ODL scenarios. The novelties and contributions of this paper can be summarized as follows.

- Fine-grained allocating of resources for the fluctuating number of participants while satisfying the end-to-end delay and minimizing the total cost (i.e. network costs and servers' costs).
- Analyzing the proposed resource allocation mechanism by modeling it as an optimization problem.
- Designing a heuristic to reach the sub-optimal solution for the large-scale scenarios in an acceptable time.

The results show the impact of the number of participants and their geographical distributions on servers' resource cost and network cost. Moreover, they show the impact of participants' geographical distribution on the required compression rate for video streams.

The rest of this paper is organized as follows. In Section II, we discuss the basics of media handling services, the requirements of these services as well as the related work. Section III presents the system model. The CRAM heuristic is presented in Section IV followed by the performance evaluation in Section V. We conclude the paper in Section VI with contributions and future research directions.

## II. BASICS OF MEDIA HANDLING SERVICES, REQUIREMENTS, AND RELATED WORK

This section briefly introduces the basics of media handling services, followed by the requirements and the review of the related work.

### A. Basics of Media Handling Services

The key media handling services used by conferencing are compressing, mixing, and transcoding. A compressor takes as input a media stream and gives as output the same media, but encoded on fewer bits. A mixer takes as input several media streams and merges them into a single media output. When it comes to a transcoder, it changes the format of the media stream it receives as input and gives as output the same stream, but in a different format.

An end-to-end media handling procedure might involve several compressors, mixers, and transcoders. The sequence in which compressors, mixers, and transcoders are composed is critical because it does have an impact on QoS and total cost.

Fig. 2 illustrates an example of three composition possibilities for an end-to-end media handling procedure with different total costs and end-to-end delays. While the total cost encompasses servers and network resources, we here only shed light on servers' resources, for illustrative purposes. Each media handling service requires an amount of resources formed of two components: a fixed component and a variable component. The fixed component represents the operational cost of the VM (e.g., OS required resources). The variable component depends on the number of media streams taken as input. Also, we assume that having more instances of a media handling service helps to reduce the service response time. Fig. 2(a) handles all mixing requests with one mixer. Therefore, it uses minimum resources on a server and minimum servers' resource cost. However, it leads to an increase in mixing and transmission time. In contrast, in Fig. 2(b), more mixers are composed. Therefore, the mixing time is reduced. But it leads to using more resources on the server(s) and higher servers' resource cost. Also, there is no improvement in video transmission time. In fig. 2(c), the transmission time is reduced by adding a compressor to the composition. However, since it uses more instances of media handling services in comparison with Fig. 2(a), the cost of the servers' resource is increased.

### B. Requirements

A crucial requirement of media handling services pertains to *scalability* or, more simply, accommodating the changing number of conference participants. For example, in one study, the number of users in the World of Warcraft (WoW), one of the most famous MMOGs, fluctuates between approximately 1.5 million and 2.5 million over 10 hours [6]. Therefore, the resource allocator for media handling services should be able to dynamically scale the actual resources needed (e.g., CPU, RAM, and storage).

Besides scalability, being *cost efficient* in resource usage is another important requirement. As an example, WoW, uses more than ten thousand servers, while most of the capacities offered by these servers remain idle most of the time [6]. Therefore, to respect cost efficiency, the scaling should be *elastic*. It means that resource allocator should add and remove resources as much as demanded.

Moreover, *meeting the QoS requirements*, such as jitter, throughput, and end-to-end delay is crucial in media handling services. In our study, we consider an end-to-end delay. Based

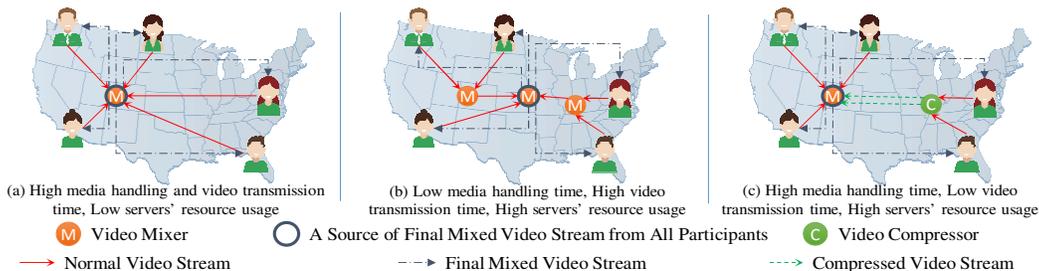

(a) High media handling and video transmission time, Low servers' resource usage
(b) Low media handling time, High video transmission time, High servers' resource usage
(c) High media handling time, Low video transmission time, High servers' resource usage

M Video Mixer    ○ A Source of Final Mixed Video Stream from All Participants    C Video Compressor
→ Normal Video Stream    -·-▶ Final Mixed Video Stream    ---▶ Compressed Video Stream

Fig. 2. Three composition possibilities of media handling services

on International Telecommunication Union (ITU), the total end-to-end delay in multimedia conferencing should not exceed 400 milliseconds [7]. This time includes the response time of all media handling services. In addition to response time, the end-to-end delay includes media transmission time. Thus, a cloud resource allocator for media handling services should consider an end-to-end delay in order to appropriately provision resources.

### C. Related Work

We categorized the related work based on the aforementioned requirements. We first review the solutions proposed so far for cloud-based multimedia conferencing. This is followed by a discussion of the other cloud-based solutions. This discussion includes multimedia solutions which are not multimedia conferencing. The solutions which are based on Network Function Virtualization (NFV) [8] are also reviewed. NFV is a technology that enables dynamic provisioning of network services. However, these solutions are discussed because NFV is also considered as a candidate technology for provisioning other services such as multimedia services [9]. Finally, the traditional approaches for multimedia conferencing are reviewed.

*1) Resource Allocation for Multimedia Conferencing in the Cloud*

Negralo *et al.* [10] present algorithms for scaling resources based on the real-time demands by using load balancing and the addition or removal of virtual machines (VMs). Reaching a predefined threshold for CPU or bandwidth usage triggers scaling. Gao *et al.* [11] also work on cost-efficient video transcoding in the cloud. They minimize the overall storage and computing cost by partially using offline and online transcoding. The focus of these works is cost efficiency and they do not consider QoS. Soltanian *et al.* in [12] consider QoS too. They propose an adaptive and dynamic scaling mechanism for cloud-based multimedia conferencing services. This work is focused on finding the optimal size of the conferencing service while considering QoS and efficiency in resource usage. However, they do not consider network cost in their model.

Hajiesmaili *et al.* [13] model the video conferencing cost in multiparty cloud video conferencing architecture. The focus of this work is minimizing the operational cost by finding the best assignment of users to VMs. They also reduce the conferencing delay. However, this work does not consider the resource allocation problem in case of having fluctuations in the number of participants. In another work, Soltanian *et al.* [14] propose a holistic cloud-based architecture for conferencing service provisioning. This work eases the provisioning of conferencing services for service and application providers. Their proposed architecture enables elastic scaling of the conferencing services. However, they do not consider the network cost in this work. Abdallah *et al.* in [15] survey other architectural works on delay-sensitive conferencing video services. They present some related applications such as Cloud Gaming and Virtual Reality (VR) and their requirements for conferencing services. In addition, they briefly talk about optimization techniques. None of the reviewed papers meet the requirement of considering fluctuations in the number of participants.

*2) Non-Conferencing Related Cloud Resource Allocation Solutions*

Several researchers have proposed solutions for resource allocation to multimedia services in the cloud. However, they do not focus on multimedia conferencing. Xavier *et al.* in [16] propose resource allocation algorithms for audio and video services in the content delivery network (CDN). The proposed solution scales the resources at the VM-level while minimizing the cost. They also consider meeting the users' quality of experience in their algorithms. Gao *et al.* in [17] present a resource allocation algorithm for transcoding as a cloud service. They maximize the service profit while achieving service processing delays. Although these works consider scalability, the elasticity and fine-grained resource allocation are not considered.

He *et al.* in [18] and Dong *et al.* in [19] consider fluctuation in the number of audio and video sources. In these two works, they consider numerous users as video broadcasters which live stream their video content such as their mobile camera feed or online game scenes. The authors in [18] propose a generic cloud framework that considers the viewers' quality of experience (QoE) and cloud resource cost. They only consider transcoding as a media handling service in their study. The authors in [19] propose an algorithm that makes a tradeoff between QoE of users and the total cost for a media service provider. None of these two works consider having a video mixing service. It means that in these works, the videos are just streamed from a source to a destination and never mixed with other video sources. There are some other cloud resource allocation works such as [20] and [21] which do not focus on multimedia services and their requirements. Therefore, these works are out of the scope of our study.

*3) NFV Resource Allocation Solutions*

There are several works done in NFV resource allocation domain. Herrera and Botero in [22] present a comprehensive survey on NFV architecture and its resource allocation problems. The reviewed works are focused on optimizing the Virtual Network Functions (VNFs) placement in the network and not focused on scaling based on the fluctuating demands.

Other researchers such as Fei *et al.* in [23] and Wang *et al.* in [24] focus on scaling the VNFs and considering the fluctuations in the demands of a service. In [23], they propose a proactive approach for provisioning VNFs by using traffic prediction. The goal of this work is to instantiate fewer VNFs to reduce cost. The authors in [24] propose an online deployment of VNF chains and dynamic scaling in response to changes in traffic. The goal of [24] is to reduce the cost by deploying a minimum number of VNFs. Also, they consider VNF placement and minimizing network congestion. The scaling in these works is in terms of a VNF instance and they do not consider increasing or decreasing the resources of existing VNFs. Dieye *et al.* in [9] introduce a cost-efficient proactive VNF placement for CDNs. In this work, the location of end-users as destinations are known in advance while the location of their surrogate servers (i.e., media sources) are not known. Similar to [23] and [24], they do not consider the elastic scaling of resources in the existing VNFs.

*4) Traditional Resource Allocation for Conferencing*

There are some resource allocation solutions for peer-to-peer (P2P) and centralized multimedia conferencing [25]. Yuen and Chan [26] reduce video transmission delay from different video sources to users. They propose an algorithm to select peers as mixers to minimize the overall delay. However, their algorithm does not account for media handling response time. Chen *et al.* [27] also propose a P2P multi-party video conferencing solution to achieve a low end-to-end delay. They optimize the streaming rates of all peers subject to network bandwidth constraints. Their study reduces the end-to-end delay without tackling the specifics of media handling services. Multipoint Control Unit (MCU) [28] is a media handling component that can include different media handling functionalities. Traditionally, all requests are handled by a single MCU, where resources are allocated in a static manner. Thus, this approach is not scalable and uses resources inefficiently.

TABLE I. Evaluation of the related work

| | | Scalability | Cost Efficiency | Elasticity | Meeting QoS |
|---|---|---|---|---|---|
| Conferencing Cloud Solutions | [10] | ✓ | ✓ | – | – |
| | [11] | ✓ | ✓ | – | – |
| | [12] | ✓ | – | ✓ | ✓ |
| | [13] | – | ✓ | ✓ | ✓ |
| | [14] | ✓ | – | ✓ | ✓ |
| Non-Conferencing Cloud Solutions | [16] | ✓ | ✓ | – | ✓ |
| | [17] | ✓ | – | – | ✓ |
| | [18] | ✓ | ✓ | – | – |
| | [19] | ✓ | ✓ | – | – |
| NFV Solutions | [23] | ✓ | ✓ | – | – |
| | [24] | ✓ | ✓ | – | – |
| | [9] | ✓ | ✓ | – | – |
| Traditional Solutions | [26] | – | – | – | ✓ |
| | [27] | – | – | – | ✓ |

Table I summarizes the evaluation of the related work with respect to the mentioned requirements of media handling services. Check marks in this table indicate that the requirement is met in the related work.

### III. SYSTEM MODEL

As it is stated in the related work, the existing works in the literature which are dealing with resource allocation for multimedia conferencing do not meet all the specified requirements. Specifically, those approaches still face efficient resource usage and scalability challenges while meeting QoS requirements. Hence, the novelty of this paper is to propose a resource allocation solution for multimedia conferencing which efficiently uses resources and scales in terms of the number of participants while guaranteeing the QoS requirements. Moreover, our solution composes different media handling services to support the participants' demands. Our system model includes the general assumptions that we made in this work and the mathematical model. In our mathematical model, we define CRAM as an Integer Linear Programming (ILP) problem.

*A. General Assumptions*

There are some assumptions that are considered to model the problem. Those categorized into two sections.

*1) Assumptions on Conferencing Applications*

We assume that conferencing applications run on a large-scale geographically distributed cloud. Also, we consider multiple conferencing participants who want to join a conferencing application and share their videos. Moreover, participants are simultaneously considered as video sources and destinations. It is assumed that the conferencing application requires the video streams from all participants to be mixed and sent to each of them.

*2) Assumptions on Media Handling Services*

Media handling services can be placed in any data center, as long as the participants' required QoS (such as latency) is satisfied. It is assumed that each media handling service is hosted on a VM. To connect media handling services, we consider different cost and latency for each network link.

Our video mixing model follows the Fork/Join parallelism technique. All video mixing requests fork off to several other mixing processes, which are executed on each video mixer until they finally join into a single mixed video. Therefore, the video mixing process for all participants depends on all video mixer instances. In this work, we assume the video mixing time in a video mixer depends on the number of input streams of that mixer. Note that all video mixers across different servers need the results from each other to complete the mixing process. Thus, the total mixing time depends on the number of video mixers and network latency.

*B. Mathematical Model*

This subsection presents our CRAM problem formulation, which is modeled as an ILP problem.

*1) Problem Statement*

Given $S$ and $U$ as sets of servers and participants (i.e., video sources and destinations) respectively, let $T_{m(k)}$ and $R_{m(k)}$ represent the time and the resource required to mix or compress $k$ video sources, respectively. We assume $T_{m(k)}$ and $R_{m(k)}$ are linear functions of $k$. Also, let $T_{a,b}$ and $P_{a,b}$ denote the time and cost to exchange a video from location $a$ to $b$, respectively. Each compressor instance can reduce the size of video by $\gamma\%$. The $T_{a,b}$ and $P_{a,b}$ are reduced by $\gamma\%$ if there is a compressor at location $a$. Also, $R_O$ are the resources which cannot be utilized for video mixing or compressing (e.g., OS required resources). There are thresholds $T_\varepsilon$ on QoS, pertaining to the maximum acceptable end-to-end delay, and $R_\varepsilon^s$ on resource capacity of server $s$. The problem is finding the minimum number of VMs and minimum network cost, while respecting QoS. Also, finding the optimal order of using media handling services to efficiently use resources is part of the problem.

We model this as an ILP problem, where we assume a media handling service to be analogous to a VM. Tables II and III delineate the inputs and variables of our problem, respectively.

*2) Objectives*

We assume the operational cost of a VM, in terms of non-utilizable resources, supersedes the cost of resources required for media handling services request of a participant (i.e. the required resources to mix or compress one new video stream in a VM), as in (1). Furthermore, we assume homogeneous costs of video mixing and compressing resources on each server. Therefore, the operational cost $R_O$, associated with a VM, inhibits the introduction of a new VM, in the event of a new participant's arrival. That is, a new VM is only instantiated if an incoming request cannot be handled by increasing the resources of an existing VM.

$$R_O \gg (R_{m\,(k+1)} - R_{m(k)}) \qquad (1)$$

TABLE II. Problem inputs

| Input | Definition |
|---|---|
| $S$ | set of servers |
| $U$ | set of users, i.e., video sources and destinations |
| $M$ | set of video mixer instances |
| $C$ | set of compressor instances |
| $V$ | set of all VMs, where $V = \{C \cup M\}$ |
| $T_{m(k)}$ | time to mix or compress $k$ video sources |
| $R_{m(k)}$ | required resources to mix or compress $k$ video sources in a VM |
| $R_O$ | non-utilizable VM operating resources |
| $T_{a,b}$ | time to send a video between location a and b |
| $P_{a,b}$ | cost to send a video between location a and b |
| $P_s$ | cost of provisioning a VM on server $s$, $s \in S$ |
| $\gamma$ | compress rate, $0 < \gamma < 100$ |
| $R_\varepsilon^s$ | threshold on the maximum amount of resources in server $s$ |
| $T_\varepsilon$ | QoS threshold (acceptable mixing response time) |
| $\beta$ | large enough constant |

TABLE III. Problem variables

| Variable | Definition | |
|---|---|---|
| $D$ | $(4\|U\|-2) \times (4\|U\|-2)$ binary matrix, where | $d_{a,b} = \begin{cases} 1, if\ 'a'is\ directly \\ connected\ to\ destination\ 'b' \\ 0, otherwise \end{cases}$ |
| $E$ | $\|U\| \times (3\|U\|-2)$ binary matrix, where | $e_{a,b} = \begin{cases} 1, if\ user\ 'a'\ directly\ or\ indirectly \\ is\ connected\ to\ VM\ 'b' \\ 0, otherwise \end{cases}$ |
| $X$ | $\|S\| \times (3\|U\|-2)$ binary matrix, where | $x_{s,v} = \begin{cases} 1, if\ server\ 's'\ hosts\ VM\ 'v' \\ 0, otherwise \end{cases}$ |
| $Y$ | $\|U\| \times (3\|U\|-2)$ matrix where, $y_{a,b}$ is the required time to transmit a video stream from user $'a'$ to VM $'b'$ and the total required time for media handling services to reach location $'b'$ | |
| $Z$ | $\|S\| \times (3\|U\|-2)$ matrix, where | $z_{s,v} = \begin{cases} g_v, if\ server\ 's'\ hosts\ VM\ 'v' \\ 0, otherwise \end{cases}$ |
| $G$ | A vector where $g_v$ is the number of users connected to the VM $v$ | |
| $F$ | $\|U\| \times (3\|U\|-2) \times (3\|U\|-2)$ binary matrix, where | $f_{i,v}^u = \begin{cases} 1, if\ user\ 'u'is\ indirectly \\ connected\ to\ VM'v'through\ VM\ 'i' \\ 0, otherwise \end{cases}$ |

Equation (2) depicts our objectives which are aiming at minimizing the overall cost. We aim to minimize the cost of allocated resources by minimizing the number of VMs. Moreover, we want to minimize the network cost. We use $x_{s,v}$ to represent a VM $v$ which is hosting on server $s$. Also, $d_{a,b}$ represents a video stream connection from source $a$ to the location $b$. The network cost between location $a$ and $b$ is shown by $P_{a,b}$.

$$\min \left\{ \sum_{s \in S} \sum_{v \in V} x_{s,v} \times P_s + \sum_{a \in U \cup V} \sum_{b \in U \cup V} d_{a,b} \times P_{a,b} \right\} \quad (2)$$

In this work, we assume the cost of sending a video from one location to another location in both directions are the same (i.e., $P_{a,b} = P_{b,a}$). Note that we know the locations of participants and servers. Therefore, to find the cost of sending a video from a participant to a VM, or from a VM to another VM, we use equations (3) and (4).

$$P_{u,v} = P_{v,u} = \sum_{s \in S} (x_{s,v} \times P_{s,u}) \quad \begin{array}{l} \forall\ v \in V \\ \forall\ u \in U \end{array} \quad (3)$$

$$P_{v_1,v_2} = \sum_{s_1 \in S} \sum_{s_2 \in S} (x_{s_1,v_1} \times x_{s_2,v_2} \times P_{s_1,s_2}) \quad \forall\ v_1, v_2 \in V \quad (4)$$

Since equation (4) is not linear, we linearize it through equations (4-1) and (4-4). We use a binary auxiliary variable $j_{s_1,s_2}$ for linearizing this equation.

$$j_{s_1,s_2} \leq x_{s_1,v_1} \quad \begin{array}{l} \forall\ v_1 \in V \\ \forall\ s_1, s_2 \in S \end{array} \quad (4\text{-}1)$$

$$j_{s_1,s_2} \leq x_{s_2,v_2} \quad \begin{array}{l} \forall\ v_2 \in V \\ \forall\ s_1, s_2 \in S \end{array} \quad (4\text{-}2)$$

$$j_{s_1,s_2} \geq x_{s_1,v_1} + x_{s_2,v_2} - 1 \quad \begin{array}{l} \forall\ v_1, v_2 \in V \\ \forall\ s_1, s_2 \in S \end{array} \quad (4\text{-}3)$$

$$P_{v_1,v_2} = \sum_{s_1 \in S} \sum_{s_2 \in S} (j_{s_1,s_2} \times P_{s_1,s_2}) \quad \forall\ v_1, v_2 \in V \quad (4\text{-}4)$$

*3) Constraints*

Based on the set $U$, we can define two sets for video mixers ($M$) and compressors ($C$). We know that each video mixer has at least two video streams as input. Therefore, set $M$ can be defined such that $|M| = |U| - 1$ and $M = \{m_1, m_2, ..., m_{|U|-1}\}$. Also, we assume we can have compressors between participants and mixers as well as between mixers. Therefore, set $C$ can be defined such that $|C| = |2U| - 1$ and $C = \{c_1, c_2, ..., c_{|2U|-1}\}$. Since each VM hosts just one media handling service, we define a set for all possible virtual machines as $V$ where $V = \{C \cup M\}$. These sets are used in the following equations.

We consider each participant has only one directed connection for sending the video stream and receiving the mixed video. Equations (5) and (6) ensure that there is only one directed connection from participants to VMs, and from VMs to participants, respectively.

$$\sum_{v \in V} d_{u,v} = 1 \quad \forall\ u \in U \quad (5)$$

$$\sum_{v \in V} d_{v,u} = 1 \quad \forall\ u \in U \quad (6)$$

Note that $d_{a,b}$ is a directed connection where $a$ and b are the head and tail, respectively. Moreover, participants need the mixed video from all others in the conference. Therefore, there is no direct connection between the participants. Equation (7) ensures this constraint.

$$\sum_{i \in U} \sum_{j \in U} d_{i,j} = 0 \tag{7}$$

To complete the video mixing process, there should be at least one VM, which is the tail of a direct or indirect connection to all original sources of video streams (i.e., participants). After finishing the whole video mixing process, the final mixed video stream should be sent to the participants from the mixers or compressors that have the whole mixing result. Equations (8) and (9) find the direct and indirect connection between all participants and all VMs. Equation (10) ensures that there is no indirect connection to any VM which has no direct connection. In addition, equations (11) and (12) consider all possible indirect connections from a participant $u$ to the VM $v$ through all other VMs. Based on these connections, equation (13) ensures that the final video streams come from the VMs which are directly or indirectly connected to all participants. Note that $e_{a,b}$ is an indirect connection where $a$ and $b$ are the head and tail, respectively.

$$e_{u,v} \geq d_{i,v} + e_{u,i} - 1 \quad \begin{array}{l} \forall\, i, v \in V \\ \forall\, u \in U \end{array} \tag{8}$$

$$e_{u,v} \geq d_{u,v} \quad \begin{array}{l} \forall\, v \in V \\ \forall\, u \in U \end{array} \tag{9}$$

$$e_{u,v} \leq \sum_{k \in U \cup V} d_{k,v} \quad \begin{array}{l} \forall\, v \in V \\ \forall\, u \in U \end{array} \tag{10}$$

$$e_{u,v} \leq \sum_{i \in V} f^{u}_{i,v} + d_{u,v} \quad \begin{array}{l} \forall\, v \in V \\ \forall\, u \in U \end{array} \tag{11}$$

$$f^{u}_{i,v} \leq \frac{d_{i,v} + e_{u,i}}{2} \quad \begin{array}{l} \forall\, i, v \in V \\ \forall\, u \in U \end{array} \tag{12}$$

$$d_{v,u} \leq \frac{\sum_{p \in U} e_{p,v}}{|U|} \quad \begin{array}{l} \forall\, v \in V \\ \forall\, u \in U \end{array} \tag{13}$$

The compressors can just reduce the video size. Therefore, the total number of input and output streams are the same. This constraint is considered in equation (14). In addition, compressors can help to reduce the size of video and in consequence, reduce the network cost and transmission time. In this work, we assume there is no need to have two consecutive compressors. Thus, there is no direct connection between the two compressor instances. Equation (15) ensures this constraint.

$$\sum_{k \in U \cup V} d_{k,c} = \sum_{k \in U \cup V} d_{c,k} \quad \forall\, c \in C \tag{14}$$

$$\sum_{i \in C} \sum_{j \in C} d_{i,j} = 0 \tag{15}$$

On the other hand, mixers are responsible to mix video streams. Therefore, at least one video mixer should be directly or indirectly connected to all participants as the tail. This constraint is ensured in equation (16).

$$\sum_{m \in M} \left\lfloor \frac{\sum_{u \in U} e_{u,m}}{|U|} \right\rfloor \geq 1 \tag{16}$$

We linearize equation (16) through equations (16-1) and (16-2) by using $h_m$ as an auxiliary variable.

$$\sum_{m \in M} h_m \geq 1 \tag{16-1}$$

$$h_m \leq \frac{\sum_{u \in U} e_{u,m}}{|U|} \quad \forall\, m \in M \tag{16.2}$$

A VM, that is hosting a media handling service, cannot be split across multiple servers. Equation (17) ensures that a VM exists on a single server. Furthermore, if there are any input streams connected to a VM, that VM should exist on one server, as depicted in (18) and (19). Also, if there are any output streams from a VM, that VM needs to exist on a server as shown in (20) and (21). Note that $\beta$ is a big enough constant used for linearization purpose.

$$\sum_{s \in S} x_{s,v} \leq 1 \quad \forall\, v \in V \tag{17}$$

$$\sum_{k \in U \cup V} d_{k,v} \leq \beta \times \left( \sum_{s \in S} x_{s,v} \right) \quad \forall\, v \in V \tag{18}$$

$$\sum_{k \in U \cup V} d_{k,v} \geq \sum_{s \in S} x_{s,v} \quad \forall\, v \in V \tag{19}$$

$$\sum_{k \in U \cup V} d_{v,k} \leq \beta \times \left( \sum_{s \in S} x_{s,v} \right) \quad \forall\, v \in V \tag{20}$$

$$\sum_{k \in U \cup V} d_{v,k} \geq \sum_{s \in S} x_{s,v} \quad \forall\, v \in V \tag{21}$$

The number of VMs and their resources are bounded by the servers' capacities. Equation (22) ensures that the required resources for media handling services and operating system in VMs are bounded by the server resource capacity.

$$R_O \times \left( \sum_{v \in V} x_{s,v} \right) + R_{m(\sum_{v \in V}(x_{s,v} \times \sum_{k \in U \cup V} d_{k,v}))} \leq R^{s}_{\varepsilon} \tag{22}$$
$$\forall\, s \in S$$

We linearize (22) by replacing it with constraints (22-1) - (22-6).

$$\sum_{k \in U \cup V} d_{k,v} = g_v \quad \forall\, v \in V \tag{22-1}$$

$$z_{s,v} \leq |U| \times x_{s,v} \quad \forall\, s \in S, \forall\, v \in V \tag{22-2}$$

$$z_{s,v} \leq g_v \quad \forall\, s \in S, \forall\, v \in V \tag{22-3}$$

$$z_{s,v} \geq g_v - |U| \times (1 - x_{s,v}) \quad \forall\, s \in S, \forall\, v \in V \tag{22-4}$$

$$z_{s,v} \geq 0 \quad \forall\, s \in S, \forall\, v \in V \tag{22-5}$$

$$R_O \times \left( \sum_{v \in V} x_{s,v} \right) + R_{m(\sum_{v \in V} z_{s,v})} \leq R_{\varepsilon} \quad \forall\, s \in S \tag{22-6}$$

The whole mixing procedure time, depends on the video mixing, compressing, and the time required for video transmission over the network. To satisfy the QoS requirement, the mixing procedure time for all participants should be less than or equal to $T_\varepsilon$. Equations (23) to (25) ensure that this end-to-end time for all participants, abides by the QoS threshold $T_\varepsilon$.

$$y_{u,v} \geq d_{i,v} \times T_{i,v} + y_{u,i} + T_{m(\sum_{k \in U \cup V} d_{k,v})} \quad \begin{array}{l} \forall\, i, v \in V \\ \forall\, u \in U \end{array} \tag{23}$$

$$y_{u,v} \geq d_{u,v} \times T_{u,v} \quad \begin{array}{l} \forall\, v \in V \\ \forall\, u \in U \end{array} \tag{24}$$

$$y_{u,v} + d_{v,u} \times T_{v,u} \leq T_\varepsilon \quad \begin{array}{l} \forall\, v \in V \\ \forall\, u \in U \end{array} \tag{25}$$

We assume that the required time for sending a video from one location to another location in both directions are the same (i.e., $T_{a,b} = T_{b,a}$). To find the time of sending a video from a participant to a VM, or from a VM to another VM, we use equations (26) and (27).

$$T_{u,v} = T_{v,u} = \sum_{s \in S}(x_{s,v} \times T_{s,u}) \qquad \begin{array}{l}\forall\, v \in V \\ \forall\, u \in U\end{array} \quad (26)$$

$$T_{v_1,v_2} = \sum_{s_1 \in S}\sum_{s_2 \in S}(x_{s_1,v_1} \times x_{s_2,v_2} \times T_{s_1,s_2}) \;\; \forall\, v_1, v_2 \in V \quad (27)$$

Since equation (27) is not linear, we use the same canonical approach used in equation (4) to linearize it.

## IV. CRAM HEURISTIC

As illustrated in Fig. 1 conferencing applications rely on a conferencing service that is offered as a SaaS. Media handling services offered as SaaS, require actual resources (e.g., CPU, RAM, and storage), that are provided by geographically distributed IaaSs. The resource allocation is performed at IaaS level using CRAM algorithm.

CRAM allows determining the number of VMs for mixers and compressors needed in order to serve a set of media handling requests. In addition, it identifies the servers that will host these VMs, together with the resulting service composition. These aspects are covered with the objective of minimizing the overall costs while meeting QoS thresholds for multimedia conferencing applications. Note that to reach the lower media handling processing time, CRAM always assigns video streams to the VMs which have fewer connected streams on each server. Also, to respect the QoS threshold, CRAM may decide for using compressors. Note that using compressors leads to lower video resolution. However, in a dense network or when participants are very far from each other, it may help to abide by the latency threshold.

Finding the best possible servers to host VMs can be mapped to the NP-hard facility location problem [29]. Besides finding the best servers to host VMs, our problem determines the best composition of media handling services. Solving our resource allocation problem for large-scale scenarios using exact algorithms is time-consuming. Thus, we introduce a heuristic to solve the problem efficiently and in a reasonable time. In this section, we propose the CRAM heuristic. It handles the composition of media handling services, together with the placement of the corresponding VMs.

The CRAM heuristic first calculates the minimum required number of VMs for mixing all streams, regardless of participants' locations. Then, it finds the possible servers with the minimum distance from all participants to host the mixers. Using these servers results in minimizing network latency and network cost. The CRAM heuristic also ensures that the available resources on these servers are enough to instantiate new VMs. Then, it checks the possibility of satisfying QoS requirements by having this minimum number of VMs hosting the mixers. If the QoS is not satisfied, the heuristic tries to increase the number of mixers (to reduce the mixing time) or add compressors (to reduce the transmission time). In these processes, our CRAM heuristic considers minimizing the cost as the main objective as well. Our solution is divided into four parts as described in Algorithms 1 to 4. We consider the constants and variables shown in Table II and Table III as the input to these algorithms. Also, to simplify the code, we assume the same resource capacity for all servers (i.e., $R_\varepsilon$).

Algorithm 1 is the main body of the CRAM heuristic. It takes as main inputs: (i) the list of participants and their locations, (ii) the list of servers and their locations, and (iii) the network transmission time and cost between different locations. This algorithm in collaboration with algorithms 2 to 4, finds the list of mixers, compressors, network connections, and the maximum end-to-end delay. This algorithm runs at the starting point of the conferencing application. In addition, it re-runs periodically to scale the system based on the fluctuations in the number of participants.

**Algorithm 1.** Media Handling Resource Allocation

**Input:**
$U, S$; // the sets of participants' and servers' locations, respectively
$P_s$ // cost of resources on a server
$R_{m(k)}, T_{m(k)}, R_O$;
$R_\varepsilon$; // the maximum capacity for all servers
$R \leftarrow R_\varepsilon$; // the set of available resources on each server
$T_\varepsilon$; // the maximum acceptable end-to-end delay
**Output:** $M, C, D$; // list of Mixers ($M$), Compressors ($C$) and the connections between participants/mixers/compressors ($D$)
$total\_delay$; // maximum end-to-end delay

*Phase 1: Find the minimum number of mixers*
1. $Min\_mixer \leftarrow 0$;
2. $handling\_time \leftarrow \infty$;
3. **do**
4. $\quad Min\_mixer \leftarrow Min\_mixer + 1$;
5. $\quad Max\_user = \mathbf{ceil}\left[\frac{|U|}{Min\_mixer}\right]$;
6. $\quad$ **If** ($handling\_time < T_{M(Max\_user)} + T_{M(Min\_mixer)}$) **Then**
7. $\quad\quad$ **return** null; //there is no possible solution for the given |U|
8. $\quad$ **end if**
9. $\quad handling\_time \leftarrow T_{M(Max\_user)} + T_{M(Min\_mixer)}$
10. **while** (($handling\_time \geq T_\varepsilon$) OR ($R_O + R_{m(Max\_user)} > R_\varepsilon$))

*Phase 2: Select the best servers for hosting mixers*
11. $vm \leftarrow 0$;
12. $i \leftarrow 0$;
13. $S \leftarrow \mathbf{DSort}(S, U)$;// sort servers based on minimum distance to the group of participants
14. **do**
15. $\quad i \leftarrow i + 1$;
16. $\quad$ **while** ($R[S[i]] \geq R_O + R_{m(Max_{user})}$ **AND** $vm < Min\_mixer$) **do**
17. $\quad\quad M[S[i]] \leftarrow M[S[i]] + 1$;// number of mixers hosted on server $i$
18. $\quad\quad vm++$;
19. $\quad\quad R[S[i]] \leftarrow R[S[i]] - (R_O + R_{m(Max\_user)})$;
20. $\quad$ **end while**
21. $\quad$ **If** ($i == |S|$ **AND** $vm < Min\_mixer$) **Then**
22. $\quad\quad$ **return** null; //not enough resources to support |U|
23. $\quad$ **end if**

24. **while** ($vm < Min\_mixer$)

*Phase 3: Check the need of compressor between mixers*

25. $used\_servers \leftarrow i$;
26. **For** j =1 $\rightarrow used\_servers$ **do**
27.   $mix\_time[S[j]] \leftarrow 0$; // maximum mixing time for each server
28.   **For** n =1 $\rightarrow used\_servers$ **do**
29.     $total\_time \leftarrow T_{M(Max\_user)} + T_{M(M[S[j]])} + T_{M(used\_servers)} + T[S[j]][S[n]]$;
30.     **if** ($total\_time \geq T_\varepsilon$) **Then**
31.       $t \leftarrow total\_time - T_\varepsilon$; // required time to compress
      //Create/assign a compressor between servers j and n
32.       $compress\_results \leftarrow Compress(j, S[n], t, "server")$;
33.       **if** ($compress\_results == Null$) **Then**
34.         **return null;** // there is no possible solution
35.       **end if**
36.       $total\_time \leftarrow T_\varepsilon$;
37.     **end if**
38.     **If** ($total\_time > mix\_time[S[j]]$) **Then**
39.       $mix\_time[S[j]] \leftarrow total\_time$; // keep track of mixing time and network transmission time between all mixers
40.     **end if**
41.   **end for**
42. **end for**

*Phase 4: Assign participants to mixers AND check the need of compressors*

43. $max\_delay \leftarrow 0$;
44. **For** u =1 $\rightarrow |U|$ **do**
45.   $total\_time \leftarrow 0$;
  //find the closest server with a mixer that can accept a participant
46.   $s \leftarrow ACS(u, M)$; //acceptable closest server to the participant $u$
47.   $total\_time \leftarrow mix\_time[S[s]] + 2 \times T[U[u]][S[s]]$;
48.   **If** ($total\_time \leq T_\varepsilon$) **Then**
    //Assign the participant $u$ to a mixer on server $s, (s \in S)$
49.     $D[u][S[s]] \leftarrow 1$; //connection from participant to server
50.     $D[S[s]][u] \leftarrow 1$; //connection from server to participant
51.   **end if**
52.   **Else**
    //Create/assign a compressor between participant u and server s
53.     $t \leftarrow total_{time} - T_\varepsilon$; // required time to compress
54.     $compress\_result \leftarrow Compress(u, S[s], t, "user")$;
55.     **if** ($compress\_results == Null$) **Then**
56.       **return null;** // there is no possible solution
57.     **end if**
58.     $total\_time \leftarrow T_\varepsilon$;
59.   **end else**
60.   **If** ($total\_time > max\_delay$) **Then**
61.     $max\_delay \leftarrow total\_time$;//maximum end-to-end delay
62.   **end if**
63. **end for**

**Return** $M, C, D, max\_delay$

Algorithm 1 has four main phases. In the first phase, it finds the minimum possible number of mixers that can mix the total number of video streams from all participants. To find this minimum number, it considers both the QoS threshold and the available resources on the servers.

After finding the minimum number of mixers, in phase two, it places these mixers on the servers which are closer to most participants. Also, it makes sure that the selected server has enough resources to host VMs. To find the servers based on the minimum distances to most participants, it uses Algorithm 2 (i.e., DSort).

After placing the mixers on the chosen servers, in phase three it checks the need of having compressors between mixers. If the total time of the mixing process and the network transmission time between two servers cannot abide by the QoS threshold, a compressor will be added between these servers. To assign or create a compressor between two servers, Algorithm 3 (i.e., Compress) is used in this phase. At the end of phase three, all mixers and required compressors between them are placed. Moreover, the mixing time for each specific server will be known.

In the last phase, participants are assigned to the closest mixer which can accept a new participant. The acceptable closest server is retrieved by using Algorithm 4 (i.e., ACS). Moreover, if the end-to-end delay is greater than the QoS threshold, it uses Algorithm 3 to assign a compressor between participant and mixer.

**Algorithm 2. (DSort)**: Sort servers based on minimum distance to a group of participants

**Input:**
$S$; // the sets of servers' locations
$U$; // the sets of participants' locations
**Output:** $Server$ // sorted list of servers

1. $delay[]$; // an array to keep track of distance for each server
2. **For** n =1 $\rightarrow |S|$ **do**
3.   **For** u =1 $\rightarrow |U|$ **do**
4.     $delay[n] \leftarrow delay[n] + T[User[u]][S[s]]$;
5.   **end for**
6. **end for**
7. $delay2[] \leftarrow sort(delay[])$;// keep sorted distances in another array
8. **For** $i = 1 \rightarrow |S|$ **do**
9.   **For** $j = 1 \rightarrow |S|$ **do**
10.     **if** ($delay2[i] == delay[j]$) **Then**
11.       $Server[i] \leftarrow j$; // keep track of server n's location
12.       $delay[j] \leftarrow -1$; //change to a negative value to make sure not using the same server more than once
13.       **break;**
14.     **end if**
15.   **end for**
16. **end for**

**Return** $Server$

Algorithm 2 sorts the servers based on their minimum distances to a group of participants. It takes the list of servers and participants and returns a list of sorted servers. This algorithm calculates the total distance from each server to all participants and uses a simple sort function (e.g., binary sort).

**Algorithm 3. (Compress):** Create or assign a compressor

**Input:** $sender$ // video sender (i.e., a participant or a server)
$b$ // the location of destination server
$t$ // minimum time that needs to be reduced by compression
$string$ // to find video sender is a participant or another server
$P$ // the matrix of video transmission costs over the network
$T$ // the matrix of video transmission times over the network
$Rate_{max}$ // the maximum acceptable compression rate (0 to 1)
**Output:** $C, D$ // list of compressors and their connections
$Rate$ // compression rate for the requested compress

1. **if** ($string == $ "server") **Then**
2.    $a \leftarrow S[sender]$; // keep location of the server in $a$
3. **else**
4.    $a \leftarrow Users[sender]$; keep location of the participant in $a$
5. **end if/else**
6. $Max\_distance \leftarrow T[a][b] - t - T_{m(1)}$;
7. $possible\_servers[]$; list of possible servers that can host compressors between locations a and b
8. $flag[] \leftarrow 0$; // to keep the demand for adding a new compressor
9. $j \leftarrow 0$;

*Phase 1: Find possible servers to host compressors between $a$ and $b$*

10. **For** $i = 1 \rightarrow |S|$ **do**
11.    **if** ($T[a][S[i]] < Max\_distance$ **AND** $R[S[i]] > R_{m(1)}$) **Then**
12.      $j \leftarrow j + 1$;
13.      $possible\_servers[j] \leftarrow i$; keep server $i$ as a possible server
14.    **end if**
15. **end for**
16. **if** ($|possible\_servers| == 0$) **Then**
17.    **return null**; // there is no possible server to host compressors
18. **end if**

*Phase 2: Find the corresponding cost for hosting or using compressors on each possible server found*

19. $C\_R \leftarrow 1 - Rate_{max}$; //
20. **For** $i = 1 \rightarrow |possible\_servers|$ **do**
21.    $s \leftarrow possible\_servers[i]$
22.    **if** ($C[S[s]] == 0$) **Then** //no existing compressor on server $s$
23.      **if** ($R[S[s]] < R_O + R_{m(1)}$) **Then** //not enough resources
24.         $Cost[S[s]] \leftarrow \infty$;
25.         continue;
26.      **end if**
27.      $Cost[S[s]] \leftarrow P[a][S[s]] + (R_{m(1)} + R_O) \times P_s$;
28.      $flag[S[s]] \leftarrow 1$;
29.    **end if**
30.    **Else**
31.      $min\_stream \leftarrow \infty$;
32.      **For** $c = 1 \rightarrow C[S[s]]$ **do**
33.         $m \leftarrow comp\_connections[S[s]][c]$; //connected number of streams to the compressor $c$ on server $s$
34.         **if** ($m < min\_stream$) **Then**
35.            $min\_stream \leftarrow m$;
36.         **end if**
37.      **end for**
38.      **if** ($T_{m(min\_stream+1)} + T[a][S[s]] + T[S[s]][b] \times C\_R \leq T[a][b] - t$) **Then**
39.         $Cost[S[s]] \leftarrow P[a][S[s]] + (R_{m(1)}) \times P_s$;
40.      **end if**
41.      **Else**
42.         **if** ($R[S[s]] < R_O + R_{m(1)}$) **Then** //not enough resources
43.            $Cost[S[s]] \leftarrow \infty$;
44.            continue;
45.         **end if**
46.         $Cost[S[s]] \leftarrow P[a][S[s]] + (R_{m(1)} + R_O) \times P_s$;
47.         $flag[S[s]] \leftarrow 1$;
48.      **end else**
49.    **end else**
50. **end for**

*Phase 3: Assign a compressor between locations $a$ and $b$ based on cost*

51. $Cost2[] \leftarrow sort(Cost[])$; // keep sorted cost in another array
52. **For** $j = 1 \rightarrow |Cost|$ **do**
53.    **if** ($Cost2[1] == Cost[S[j]]$) **Then**
54.      $chose \leftarrow possible\_server[j]$; // chosen server to host the compressor between $a$ and $b$
55.      **break**;
56.    **end if**
57. **end for**
58. **if**($string == user$)**Then**
59. $D[sender][S[chose]] \leftarrow 1$; connection from sender to server
60. $D[S[chose]][sender] \leftarrow 1$; connection from server to sender
61. **else**
62. $D[S[sender]][S[chose]] \leftarrow D[S[sender]][S[chose]] + 1$;
63. **end if/else**
64. $D[S[chose]][b] \leftarrow D[S[chose]][b] + 1$;
65. $C[S[chose]] \leftarrow C[S[chose]] + flag[S[chose]]$;
66. $min\_stream \leftarrow \infty$;
67. $used\_compressor \leftarrow 0$;
68. **For** $c = 1 \rightarrow C[chose]$ **do**
69.    $m \leftarrow comp\_connections[S[chose]][c]$; // number of streams
70.    **if** ($m < min\_stream$) **Then**
71.      $min\_stream \leftarrow m$;
72.      $used\_compressor \leftarrow c$;
73.    **end if**
74. **end for**
75. $comp\_connections[S[chose]][used\_compressor] \leftarrow comp\_connections[S[chose]][used\_compressor] + 1$;

*Phase 4: Find the required compression rate for this stream*

76. $New\_t_{s,b} \leftarrow T[a][b] - t - T[a][S[chose]] - T_{m(min\_stream+1)}$;
77. $Real\_Rate \leftarrow (T[S[chose]][b] - New\_t_{s,b})/T[S[chose]][b]$;

**Return** $C, D, Real\_Rate$

The CRAM heuristic considers video mixing and compressing as two main media handling services. The compressing process is described in Algorithm 3. It has three main inputs: (i) two locations that need a compressor in between, (ii) the minimum time that needs to be reduced by compression, and (iii) the video mixing transmission times and costs between different locations. Our proposed compression algorithm does not have a fixed compression rate. It compresses as less as possible to have less impact on the video resolution. We also consider a maximum acceptable compression rate (i.e., $Rate_{max}$) as the input for this algorithm.

The compression algorithm has four main phases. In phase one, it finds the servers that are close enough to the video sender and have resources to compress a video stream. According to the servers found, in phase two, it calculates the corresponding cost for assigning the compressing request for each server. The cost is calculated based on the server's resource cost and the network transmissions cost. If the chosen server has no compressor on it, this phase considers the cost of creating a new compressor on the server in the total cost. However, if there is an existing compressor on the server, this phase checks if the compressor can accept another stream. It ensures by checking the satisfaction of the minimum time that needs to be reduced by compression. In the case of satisfaction, there is no extra cost for creating a new VM and the server cost is calculated based on the required resources to compress one more stream. On the other hand, if it cannot satisfy, then another compressor needs to be created on this server and the cost of a new VM will be considered.

According to the calculated cost to host a compressor for each server, phase three selects the server with the minimum cost and allocates the required resources for the compressor. Also, it creates a link from the sender to the compressor and from the compressor to the destination. If there is more than one compressor on the chosen server, it always assigns the video stream to a compressor with minimum connected streams. It helps to minimize the overall media handling time. At the end of this algorithm, in phase four it calculates the exact reduced time by compression and also finds the compression rate.

**Algorithm 4. (ACS): Find the acceptable closest server**

**Input:** $M$ // list of Mixers
$u$ // a participant
$S$ // the sets of servers' locations
**Output:** $s$ //proposed server with mixer to host $u$

*Phase 1: Find acceptable servers*

1. $j \leftarrow 0$;
2. **For** $i = 1 \rightarrow |S|$ **do**
3.   **if** ($M[S[i]] > 0$) **Then**
4.     **For** $m = 1 \rightarrow M[S[i]]$ **do**
5.       **if** ($mixer\_connections[S[i]][m] < max\_user$) **Then**
6.         $j \leftarrow j + 1$;
7.         $possible\_servers[j] \leftarrow i$; keep server $i$ as a possible server
8.         **break**;
9.       **end if**
10.     **end for**
11.   **end if**
12. **end for**

*Phase 2: Find the closest server from the acceptable servers*

13. $min\_distance \leftarrow \infty$;
14. $s \leftarrow 0$;
15. **For** $i = 1 \rightarrow |possible\_servers|$ **do**
16.   **if** ($min\_distance > T[U[u]][S[possible\_servers[i]]]$) **Then**
17.     $s \leftarrow possible\_servers[i]$; //chosen server to assign the participant to a mixer
18.     $min\_distance \leftarrow T[U[u]][S[possible\_servers[i]]]$;
19.   **end if**
20. **end for**
21. $min\_stream \leftarrow \infty$;
22. **For** $m = 1 \rightarrow M[S[s]]$ **do**
23.   **if** ($mixer\_connections[S[s]][m] < min\_stream$) **Then**
24.     $mixer \leftarrow m$; // chosen mixer to support participant
25.     $min\_stream \leftarrow mixer\_connections[S[s]][m]$;
26. **end for**
27. $mixer\_connections[S[s]][mixer] \leftarrow mixer\_connections[S[s]][mixer] + 1$;

**Return** $s$

Algorithm 4 is responsible to find the closest server which is hosting a video mixer to a participant. It has two main phases. In the first phase, it finds the servers with at least one video mixer whose total connected streams is less than a maximum possible connection calculated in phase one of Algorithm 1. In phase two, it selects the one which is closest to the participant. Also, it selects the video mixer on this server with the minimum connected streams to be responsible for this mixing request. In addition, it increases the number of connected video streams for the selected video mixer.

## V. SIMULATION RESULTS

This section describes our evaluation scenarios and the simulation settings followed by the obtained results.

### A. Evaluation Scenarios and Simulation Settings

We consider two different conferencing applications as our evaluation scenarios. (i) Massively Multiplayer Online Game (MMOG) and (ii) Online Distance Learning (ODL). In these scenarios, the conference participants are sharing their videos in the logic of the application. The aim is to allow each participant to have a mixed video from all other participants. In MMOG, participants are from different geographical locations in the world. Thus, the end-to-end delay may be high. In contrast, in ODL, the number of participants is limited, and they are distributed in a smaller area such as one country. For our simulation, we consider two different geographical distributions for participants as depicted in Fig. 3. (a) *Homogeneous* – participants are distributed over the whole area (i.e., world or country) with similar density. (b) *Heterogeneous* – the majority of the participants are geographically distributed in the east and the west side of the area. These distributions can help to understand the behavior of the proposed solution when the participants are close or far from each other.

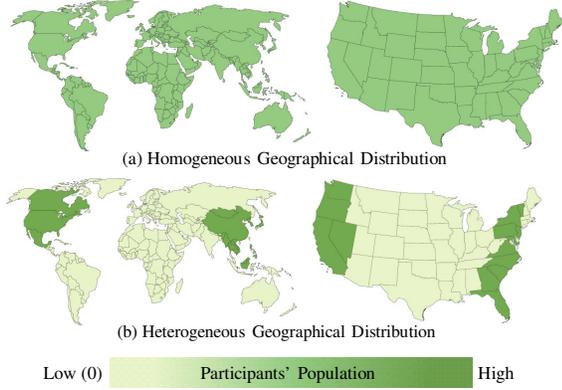

Fig. 3. Geographical distribution of participants in conferencing applications

TABLE IV. Simulation parameters and settings

|  | MMOG | ODL |
|---|---|---|
| Number of servers | 20 | 9 |
| Servers geographical distribution | Over the world | Over the USA |
| Number of participants | 100, 2000, 3000 | 100, 200, 500 |
| Participants' geographical distribution | Homogeneous: Equally distributed in each server's location | |
| | Heterogeneous: Half of users are in the western city and half are in the eastern one | |
| $T_{m(k)}$ | 6 msec per video source | |
| $R_{m(k)}$ | 20 MB (RAM) per video source | |
| $R_o$ | 400 MB (RAM) | |
| $R_\varepsilon^s$ | 10240 MB (RAM) per each server | |
| $T_\varepsilon$ | 400 msec | |
| $P_s$ | $0.01 per MB | |
| $T_{a,b}$, $P_{a,b}$ | As in [1] | |
| Maximum acceptable compression rate | 0.95 | |

For our simulations, we consider having servers in twenty cities over the world for MMOG and nine cities over the USA for ODL. For the network transmission time between servers, we use the information available at [30]. Fig. 4 shows the locations of considered servers. Also, we consider different number of participants for both scenarios. We assume a snapshot of the number of participants in this work. To study the impact of servers' resources and network costs, we consider various settings with different simulation parameters. We assume that the network transmission cost between two locations is a linear function of the transmission time between them. In fact, the farther two locations are, the higher is the network cost between them. Also, for the media handling time and required resources, we consider our prototype experience in [5]. The simulation parameters and settings are depicted in Table IV. In our evaluation, we account for the server resource in terms of used memory. However, the mathematical model and our heuristic are general enough to accommodate other types of resources as well.

*B. Results*

We solve our mathematical model to achieve optimality for the small-case scenario using LPSolve Java Library[1]. For the medium-scale and large-scale scenarios (i.e., scenarios with a higher number of participants) deriving the optimal solution with exact algorithms used by the solver is very time-consuming. Therefore, we only present the results of our heuristic that can support the number of participants in our simulation settings. However, the results in the small-case scenario allow us to validate our mathematical model. In addition, they show that our mathematical model enables the orchestration of media handling services and the possibility of composing these services on the fly. As an example of the result of the mathematical model for a small-case scenario, we ran our model while having 6 participants in Seattle and 2 participants in Toronto. The result shows a composition of one video mixer and one compressor. It allocates required resources for the video mixer in Seattle and for the compressor in Toronto.

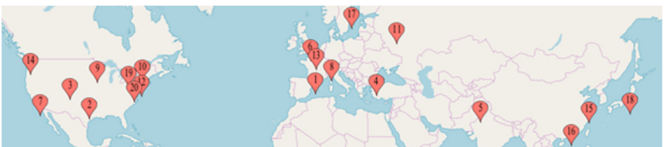

Fig. 4. Geographical distribution of the servers

In ODL, we assume all participants are from the USA with homogeneous or heterogeneous geographical distributions. We run the CRAM heuristic for 100, 200, and 500 participants. Fig. 5 shows the total cost by considering both servers' resources and network costs. By increasing the number of participants, the need for media handling services increases. This leads to allocating more resources and implies higher communication traffic as well. Thus, as depicted in fig. 5, the total cost increases as a higher number of participants is considered. However, considering the same number of participants, the total cost in homogeneous geographical distribution is greater than that of the heterogeneous geographical distribution. The reason is that the heterogeneous geographical distribution favors the execution of some media handling services locally. By that, it leads to transmit a lower number of streams over the network and implies a lower total cost.

Fig. 6 depicts the servers' resources (i.e., RAM) that is allocated for media handling services. By increasing the number of participants, our heuristic allocates more resources to media handling services to cope with the requests. The amount of memory allocation for the same number of participants is greater in the case of heterogeneous geographical distribution. In fact, in the homogeneous geographical distribution of ODL, most of the participants can reach the mixers without the need of passing through the compressors. It leads to using fewer compressors in homogeneous and less memory allocation compared to heterogeneous.

Fig. 7 shows the network cost. By increasing the number of participants, the traffic grows, implying a higher network cost.

---

[1] http://lpsolve.sourceforge.net/

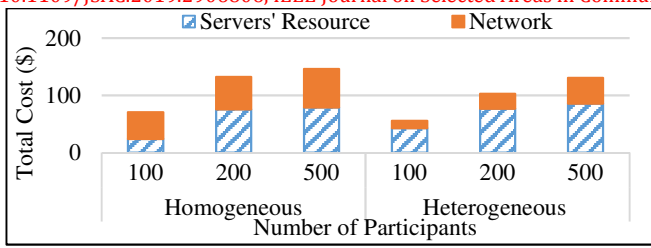

Fig. 5. CRAM heuristic total cost in ODL

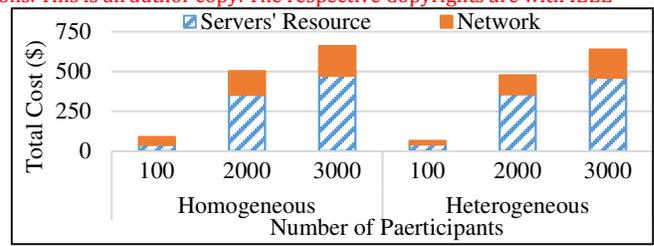

Fig. 9. CRAM heuristic total cost in MMOG

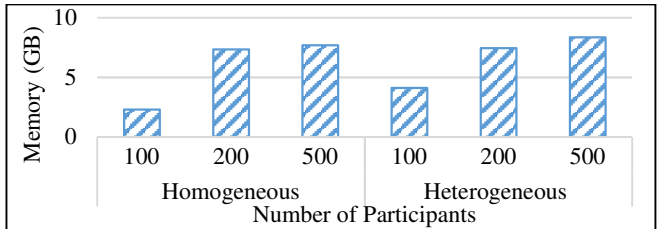

Fig. 6. CRAM heuristic total memory allocation in ODL

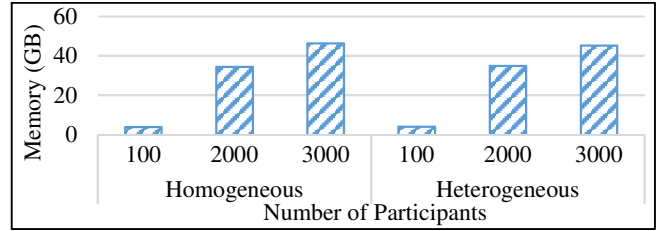

Fig. 10. CRAM heuristic total memory allocation in MMOG

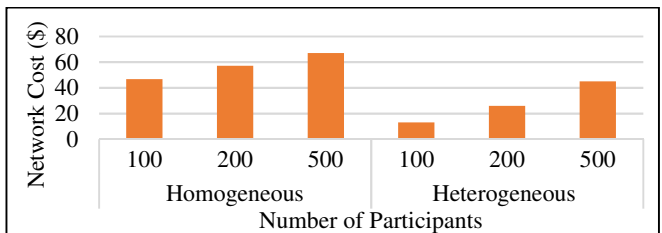

Fig. 7. CRAM heuristic network cost in ODL

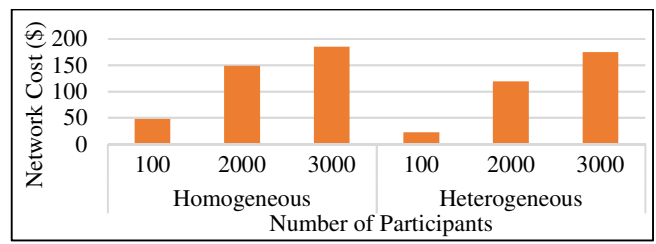

Fig. 11. CRAM heuristic network cost in MMOG

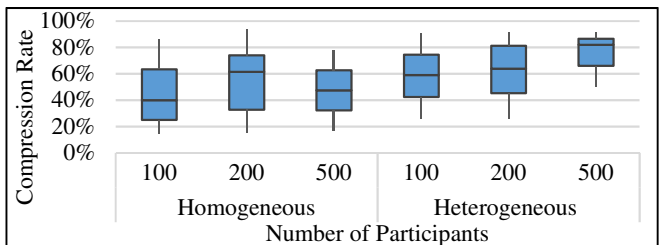

Fig. 8. CRAM heuristic video compression rate in ODL

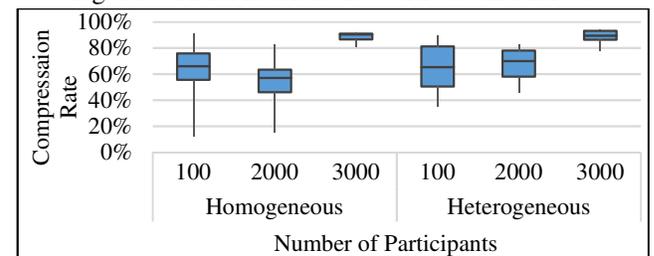

Fig. 12. CRAM heuristic video compression rate in MMOG

Unlike servers' resources, the network cost is less in heterogeneous geographical distribution in comparison with homogeneous for the same number of participants. In fact, the aggregation of participants helps to decrease network communications and reduces the network cost. However, as it is depicted in Fig. 8, it causes more compression rate in heterogeneous in comparison with homogeneous geographical distribution for the same number of participants. In fact, the compressors should serve a higher number of participants in heterogeneous geographical distribution. Thus, it increases the compression rate to cope with the QoS threshold and reduces the network transmission time. The lines in the boxes indicate the median for the compression rate.

On the other hand, in MMOG, we assume all participants are from different locations in the world. In this scenario, CRAM heuristic runs for 100, 2000, and 3000 number of participants. As depicted in Fig. 9, similar to the ODL, by increasing the number of participants, the total cost will increase as well. Also, the total cost for the same number of participants in heterogeneous geographical distribution is less than that of the homogeneous geographical distribution. Based on that, both evaluation scenarios show that regardless of the area size, the aggregation of participants can help reduce the total cost.

The memory allocations for different numbers of participants in MMOG is depicted in Fig. 10. Unlike the results of ODL, the memory allocation for MMOG in both homogeneous and heterogeneous geographical distributions are almost the same. The reason is that in MMOG, even in the homogeneous geographical distribution, the participants are far from each other. This leads to using several compressors. In fact, the aggregation of the participants into two locations does not help to reduce the required resources for compressing service. However, as it is depicted in Fig. 11, the aggregation can help to reduce the network cost in heterogeneous geographical distribution. Although the network cost is decreased by the aggregation, it leads to more compression rate as it is shown in Fig. 12. In other words, more participants end up with lower video resolution in comparison with homogeneous geographical distribution.

For the composition, the CRAM heuristic orchestrates the required instances of media handling services for participants. Note that each participant may follow a specific media handling composition which differs from others. Fig. 13 shows an example of the created compositions for two different participants in different locations. As shown in the figure, CRAM may assign the participant from Seattle to a mixer which is hosted by a server in Seattle. Thus, this participant will receive the final mixed stream from that mixer as well. However, if CRAM allocates resources to the mixers in Seattle and a participant from Toronto wants to use the mixers, to respect the maximum latency, CRAM allocates a compressor in a location which reduces the total cost and assigns the participant from Toronto to it. Then, the result of compression is sent to the mixer in Seattle. For this specific example, CRAM allocates a compressor on the Seattle server as well. Therefore, the final results are compressed one more time and then it sends to the participant in Toronto.

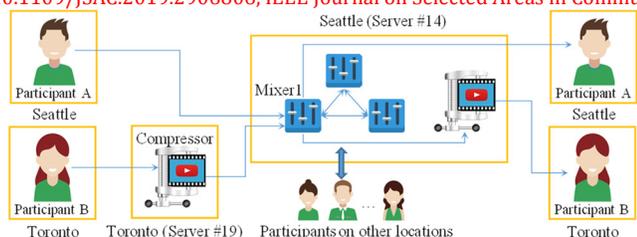

Fig. 13. Two different media handling compositions for users

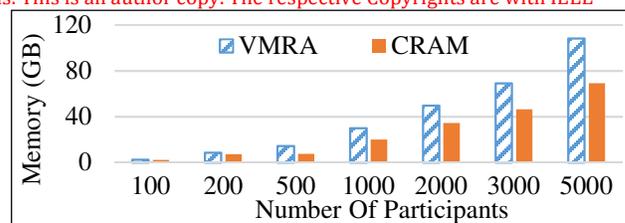

Fig. 14. CRAM vs. VMRA Servers' Resource Usage

*C. Comparison Results*

As it is described in the related work section, to the best of our knowledge, CRAM is the first work on cloud-based resource allocation for media handling services that meets all the requirements of scalability, cost-efficiency, elasticity, and meeting the QoS. However, to compare the CRAM heuristic with other algorithms, we choose VMRA [5], as the baseline. The focus of the VMRA algorithm is minimizing the servers' resource cost while meeting the QoS requirements. We use the same parameters mentioned in Table IV for both CRAM and VMRA simulations. In VMRA, servers and participants are divided into different zones. To support a higher number of participants, we consider 7 zones in the VMRA setting. In addition, since in the VMRA the network cost is not considered, we choose the servers which their distances are less than 50 milliseconds. Fig. 14 is the comparison results between CRAM's used memory and the one in VMRA. Both CRAM and VMRA can mix the video stream for the participants less than the QoS threshold. However, as it is depicted in Fig. 14, the VMRA memory usage is more than that of the CRAM for any same number of participants. The main reason is that in VMRA, video streams are mixed in each zone and the results will be sent to other zones to be mixed again. However, in CRAM, there are no zones and video mixers that can be placed anywhere based on demand. It leads to less number of video mixers and in consequence, consumes fewer resources in the CRAM. The average time to run the CRAM and VMRA heuristics are 154 and 32 milliseconds, respectively. Although the execution time of CRAM is more than VMRA, this time is negligible.

## VI. CONCLUSION

We propose a novel and scalable cloud-based resource allocation mechanism for media handling services. Our mechanism enables efficient utilization of leased resources. It allocates resources for fluctuating number of participants while meeting the end-to-end delay constraint. The proposed mechanism composes multimedia conferencing services from different media handling services to support the participants' demands. We model the problem as an ILP problem and design a heuristic to solve it over large-scale scenarios. Our simulation results show that the number of participants and their geographical distribution have a significant impact on the servers' resource cost, the network cost, and the required compression rate for video streams. In addition, the comparison results show that for the same number of participants, CRAM uses less resources than the closest algorithm in the literature. As the future work, prediction algorithms to predict participants' arrivals can be introduced. In addition, the proposed solution can be evaluated with real datasets. Moreover, instead of considering the snapshots, algorithms can be designed with finer scalability (i.e., handling the joining and leaving of participants separately).

ACKNOWLEDGMENT

This work is partially the Canada Research Chair Program, and the Canadian Natural Sciences and Engineering Research Council (NSERC) through the Discovery Grant program